# Preparation of atomically-flat SrTiO$_3$ surfaces using a deionized-water etching and thermal annealing procedure

J. G. Connell, B. J. Isaac, D. R. Strachan, and S. S. A. Seo[*]

*Department of Physics and Astronomy, University of Kentucky, Lexington, KY 40506*

We report that a deionized water etching and thermal annealing technique can be effective for preparing atomically-flat and singly-terminated surfaces of single crystalline SrTiO$_3$ substrates. After a two-step thermal-annealing and deionized-water etching procedure, topography measured by atomic force microscopy shows the evolution of substrates from a rough to step-terraced surface structure. Lateral force microscopy confirms that the atomically-flat surfaces are singly-terminated. Moreover, this technique can be used to remove excessive strontium oxide or hydroxide composites segregated on the SrTiO$_3$ surface. This acid-etchant-free technique facilitates the preparation of atomically-aligned SrTiO$_3$ substrates, which promotes studies on two-dimensional physics of complex oxide interfaces.

[*]E-mail: a.seo@uky.edu

Recently, novel electronic properties have been discovered at the interfaces of complex oxides: a high-mobility two-dimensional electron gas (2DEG) at the interface of $LaAlO_3/SrTiO_3$[1-5], which has been shown to exhibit both superconductivity[6] and ferromagnetism[7,8] as well as electronic reconstructions and superconductivity at the interfaces of $LaTiO_3/SrTiO_3$[9-12], $LaVO_3/SrTiO_3$[13], and magnetic ordering in $LaMnO_3/SrTiO_3$ superlattices[14]. Orbital reconstructions[15] and high-$T_c$ superconductivity[16] in cuprite interfaces along with magnetoelectric[17] and strain-tuning[18] effects in $LaMnO_3/SrMnO_3$ superlattices have also been observed. Studying intriguing interfacial phenomena of complex oxides provides important clues not only to understanding the physics of complex oxides but also to advancing oxide electronics[19].

Preparation of atomically-flat surfaces of substrates is an indispensable step to successfully prepare well-characterized samples and to achieve the intriguing electronic properties at the interfaces. Since the atomically-abrupt interfaces can be achieved only when substrates are flat, effective methods of preparing flat substrates cannot be overemphasized. Atomically-flat surfaces of strontium titanate ($SrTiO_3$) single crystals, which are by far the most widely used substrate in complex-oxide research, are typically achieved by using an acid-based etchant[20-22] and thermal-annealing. For example, the method based on buffered-hydrofluoric acid (BHF) etching, which is the same chemical etching procedure used in silicon semiconductor research and industry for removing $SiO_2$, resulted in atomically-flat $SrTiO_3$ substrates.[20,22-25] Owing to the safety issues of acidic etchants, these acid-based methods have been a barrier against promoting active research of novel interfacial properties of complex oxides.

In this letter, we show that a *non-acidic* deionized (DI)-water treatment and thermal annealing technique can be effectively used to prepare atomically-flat and singly-terminated surfaces of $SrTiO_3$ substrates. The perovskite $SrTiO_3$ has two possible surface terminations along the (100) direction: SrO and $TiO_2$. We note that the SrO layer of $SrTiO_3$ has an ionic bonding nature, which is in contrast with the covalent bonding of $SiO_2$. Thus water ($H_2O$) might be effective to chemically etch the SrO layer since SrO is a water-soluble material [25-27]. By combining two thermal annealing steps before and after DI-water treatment, our process successfully removes SrO terminated layers as well as what are most likely segregated strontium oxide or hydroxide islands from the surface.

We took commercially available SrTiO$_3$ (100) and SrTiO$_3$ (111) single crystalline substrates (Crystec GmbH) and prepared them by annealing at 1000°C for 1 hr. (1$^{st}$ thermal annealing) in air. Substrates were then rinsed with DI-water (Resistance > 15 MΩ) via agitation for about 30 seconds at room temperature. Finally, the substrates were annealed again at 1000°C (2$^{nd}$ thermal annealing) for 1 hr. in air. The surfaces of the substrates were characterized through the etching and annealing process using an Atomic Force Microscope (Park XE-70).

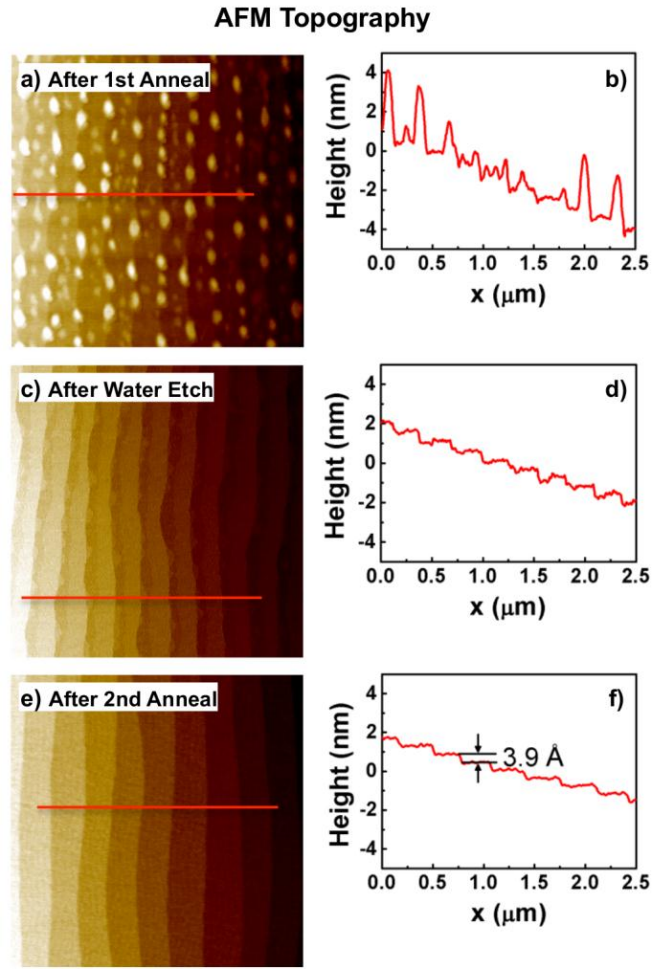

**FIG. 1**. (Color online) Surface evolution of a SrTiO$_3$ (100) substrate through the deionized-water etching and thermal-annealing process. **a)** AFM topography after 1$^{st}$ thermal annealing with line profile **b)**. **c)** AFM topography after DI-water etching with line profile **d)**. **e)** AFM topography after 2$^{nd}$ thermal annealing. As shown in **f)** the final substrate is atomically flat with rms roughness of ~0.2 Å. Scan area is 3 × 3 μm$^2$.

Figure 1 illustrates surfaces of a SrTiO$_3$ (100) substrate that is prepared by the process at each step. Figure 1a shows the sample after the 1$^{st}$ annealing step. The rather large islands formed on the sample surface are due to strontium oxide or hydroxide segregation[28]. Figure 1c supports this hypothesis as the water soluble SrO and SrO hydroxide islands have been removed by DI-water treatment. It is noteworthy that there are half-unit-cell high step-terraces as shown in Fig. 1c and 1d. Figure 1e and 1f show the sample following the 2$^{nd}$ thermal annealing. An atomically-flat surface has been produced by the process with a step height of 3.9 Å and roughness of approximately 0.2 Å.

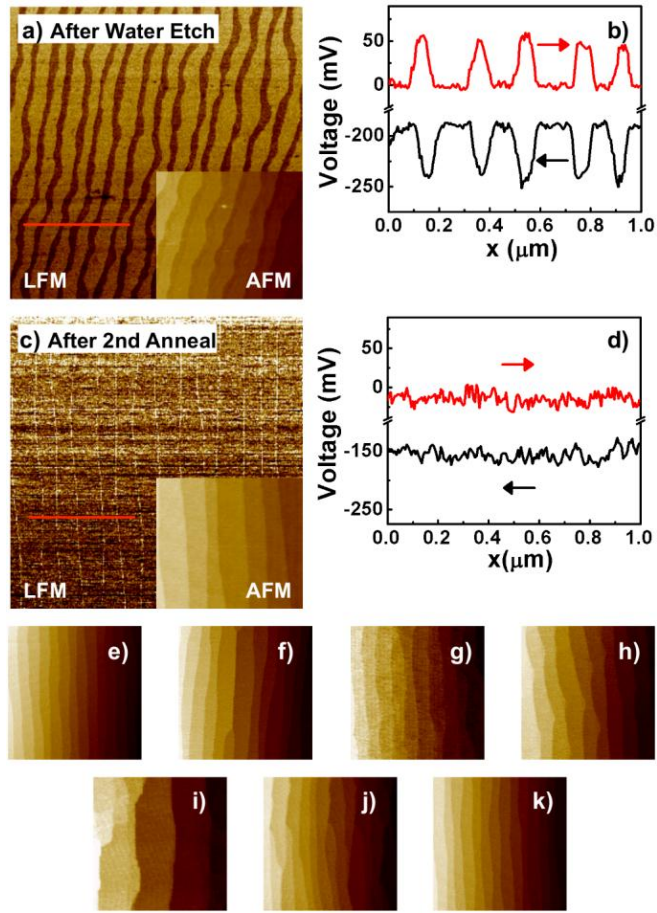

**FIG 2**: (Color online) **a)** Lateral Force Microscopy (LFM) of a SrTiO$_3$ (100) substrate after DI-water etching and **b)** the line profiles of LFM signal trace and retrace (arrows). **c)** The same substrate after 2$^{nd}$ annealing and **d)** the line profile of LFM signal trace and retrace (arrows). The insets in **a)** and **c)** show corresponding AFM topography. **e)-k)** Examples of SrTiO$_3$ (100) substrates that are tested by using the same etching and annealing process. All substrates are atomically-flat and singly-terminated. Scan area is $3 \times 3$ μm².

Clearly, an atomically-flat substrate surface has been produced. However, it is not clear just from AFM topography that this sample is singly-terminated. Thus, Lateral Force Microscopy (LFM) was employed as has been previously used on SrTiO₃ and other perovskite surfaces[28,29] to check surface-termination. Figure 2a (2b) and 2c (2d) show the LFM images (and line scans) before and after the 2$^{nd}$ thermal annealing, respectively. Note that the Figure 2a LFM image shows the regions of non-uniform friction near the step edges which correspond to the half-unit-cell deep regions in the inset. However, after a second annealing, Figure 2c and 2d show that the LFM topography has a uniform frictional response with a similar offset to the lower frictional regions in figure 2b, which implies that the substrate is singly-terminated. We tested more substrates to see whether our method is reproducible: Figure 2e – 2k display AFM topography images of a few SrTiO₃ (100) substrates which have been treated through the same DI-water etching and thermal annealing process. Atomically flat surfaces are obtained regardless of their step-terrace widths and miscut angles. Step-bunching is observed in some substrates. However, the step bunching is not due to our method but due to substrate's crystalinity since we have observed similar step-bunching in BHF treated substrates as well.

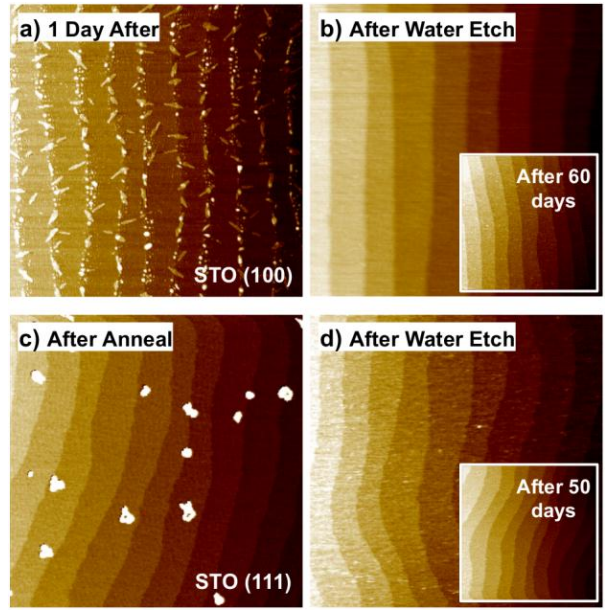

FIG 3: (Color online) AFM topography images of **a)** a SrTiO₃ (100) substrate 1 day after the second annealing and **b)** after another DI-water etching. The inset shows the same substrate 60 days later. **c)**

$SrTiO_3$ (111) substrate after the second annealing and **d)** after DI-water etching. The inset shows the same substrate 50 days later. Scan area is $3 \times 3$ μm$^2$.

We have shown that substrates that have been prepared via the DI-water etching and thermal annealing process are both atomically-flat and singly-terminated. However, surface-degradation due to SrO segregation or Sr out-diffusion has been observed in a few hours or days after the 2$^{nd}$ annealing, as shown in Figure 3a. SrO segregations are visible near the step-edges and similar surface-degradation has also been reported for BHF-etched and thermally annealed substrates[23,24]. Figure 3b shows that these degraded surfaces can be improved by another 30 sec DI-water etching step: The segregated SrO islands are removed and atomically flat surface is recovered. The inset of Figure 3b shows that the substrate preserves the surface even 60 days after the DI-water etching step. LFM confirms that the surface is still singly-terminated (data not shown). This implies that the final DI-water treatment results in rather chemically stable surfaces free from SrO segregation. We also tested our method on $SrTiO_3$ (111) substrates which have until now been typically prepared using BHF[22,30]. Figure 3c shows that $SrTiO_3$ (111) substrates also have SrO segregations on their surfaces after our process. However, after one more DI-water etching step, SrO segregations have once again been removed and the surface is clean and atomically flat as shown in Figure 3d. The inset of Figure 3d shows that the substrate surface is flat even 50 days after the treatment. Thus our method works for both $SrTiO_3$ (100) and (111) substrates.

In conclusion, atomically-flat singly-terminated surfaces of $SrTiO_3$ (100) and (111) substrates were prepared using a DI-water etching and thermal annealing technique. Furthermore, better chemically stable surfaces can be obtained when the final step in the process is DI-water etching. While preparing this manuscript, we learned of two publications where a warm water-bath was used to prepare $SrTiO_3$ substrates[31,32], which showed consistent results with ours. However, in our study we have used unheated DI-water for etching. Our *acid-etchant-free* technique eliminates the safety issues of the acid-based etching process, which will promote the progress of research on interfacial physics of complex oxides.

The authors thank G. W. J. Hassink and G. Ekanayake for useful discussions and comments. This research was supported in part by the NSF through Grant EPS-0814194 the Center for Advanced Materials, a grant from the Kentucky Science and Engineering Foundation as per Grant Agreement #KSEF-148-502-12-303 with the Kentucky Science and Technology Corporation, and a Research Support Grant from the University of Kentucky Office of the Vice President for Research.